\documentclass[a4paper,preprint]{revtex4}
\usepackage{graphicx}
\usepackage{courier}
\usepackage{amsmath}
\begin{document}

\title{
Isotopic replacement in ionic systems: the \\
$\mathrm{^4He}_2^+ + \mathrm{^3He} \longrightarrow \mathrm{^3He^4He}^+ +
\mathrm{^4He}$ reaction
\footnote{This work is affectionaly dedicated to Prof. Volker Staemmler on the
occasion of his 65th birthday: to a dear friend, an articulate scientist and a
great scholar of He$^+$-containing systems.}
}

\author{Enrico Bodo}
\affiliation{Department of Chemistry,
University of Rome La Sapienza, Piazzale A. Moro 5, 00185 Rome,
Italy}

\author{ Manuel Lara}
\thanks{Present address: JILA, University of Colorado, UBC 440,
  Boulder CO., 80309 USA}
\affiliation{Department of Chemistry,
University of Rome La Sapienza, Piazzale A. Moro 5, 00185 Rome,
Italy}

\author{Franco A. Gianturco}
\thanks{Corresponding author: Dep. of Chemistry, University of Rome {}``La
Sapienza'', P. A. Moro 5, 00185, Rome, Italy. Fax: +39-06-49913305. }
\email{fa.gianturco@caspur.it}
\affiliation{Department of Chemistry,
University of Rome La Sapienza, Piazzale A. Moro 5, 00185 Rome,
Italy}

\begin{abstract}
Full quantum dynamics calculations have been carried out for the ionic
reaction $\mathrm{^4He}_2^+ + \mathrm{^3He}$ and state-to-state
reactive probabilities have been obtained using both a time-dependent
(TD) and a time-independent (TI) approach. An accurate ab-initio
potential energy surface has been employed for the present quantum
dynamics and the two sets of results are
shown to be in agreement with each other. 
The results for zero total angular momentum suggest a marked
presence of atom exchange (isotopic replacement) reaction with
probabilities as high as 60\%. The reaction probabilities
are only weakly dependent on the initial vibrational state of the
reactants while they are slightly more sensitive to the degree of
rotational excitation. A brief discussion of the results for selected higher
total angular momentum values is also presented, while the
$l$-shifting approximation \cite{jshift} has been used to provide 
estimates of the total reaction rates for the title process. Such rates are
found to be large enough to possibly become experimentally accessible. 
\end{abstract}

\maketitle

\section{Introduction}	
The study of helium nanodroplets has been shown over recent years to
provide a very interesting and novel medium  that acts as a
"quantum matrix" in which one can probe the ro-vibrational and
electronic spectroscopy of the various dopants which can be
"solvated" in it and in which one can also observe the nonclassical
effects caused by the surrounding He atoms \cite{1,2}, together with
the changes due to their bosonic/fermionic symmetry properties
\cite{3}. The experiments which can perform ionization of the helium
matrix further raise the issue of the mechanism leading to
ionization vs fluorescence, and of the understanding of the features
of the interaction of the resulting ion with the atoms in
the droplet which can show evidence of marked
quantum many-body effects \cite{4,5}.

One of the  nanoscopic consequences of the formation of a permanent
cationic impurity within the droplet is its high mobility within the
weakly interacting solvent, whereby the positive charge can in principle
migrate within it by some resonant charge hopping mechanism
\cite{6}. This process is taken to be terminated either by formation
of a strongly bound He$_2^+$ ion or by charge transfer to a dopant
species. Our recent calculations on that process \cite{7,8} have
shown that termination can also be achieved by multiple inelastic
collisions within the droplet that enhance the
evaporation of atoms from the latter. Furthermore, possible
branching reactions of either the dopant ions or of the ionic
moiety He$_2^+$ could also occur within the
droplet because either species will trend to move towards the center of
the cluster in order to minimize the total potential energy of the
composite system \cite{9}.

Recent experiments have been carried out on specific reactions initiated by
primary ionization of the droplet \cite{10,11} and have shown 
that a rapid quenching of the internal degrees of freedom of
the partner species is favored by the superfluid environment. The
corresponding reactive branching of the ionic partners
is therefore also markedly affected and could cause, 
in some cases, stabilization of intermediate complexes different from those
expected, and seen, in the gas phase \cite{11}. In a recent
theoretical and computational study \cite{12} we have in fact shown
the possibility of having strongly exothermic chemical reactions without 
activation barrier, 
the occurrence of which should be greatly facilitated by the presence of
the inert, quantum environment at the low temperatures existing within
the helium "nanocryostat" \cite{12}.

In the present analysis we therefore wish to present  
calculations where the stabilized primary ion in the
droplet, the He$_2^+$ species occupying its lowest ro-vibrational
level, is made to react with one
of the adatoms of the environment. The idea is to analyze several
aspects of the problem which carry considerable theoretical interest
and which can also guide us to understand the possible behavior of
the ionized droplet under actual experimental conditions.

In particular, taking advantage of our recent calculations of the
potential energy surface (PES) for the He$_2^+$+He system
\cite{8}, we have studied the quantum reaction for the isotopic
exchange, i.e. for the molecule $^4$He$_2^+$ colliding with neutral
$^3$He, in order to add physical distinguishability  within the
process at hand. We have carried out the calculations, as we shall
further describe below, using both a quantum time independent (TI)
approach and a quantum, time dependent (TD) wavepacket approach.
As we shall see, both methods produce essentially the same results
and therefore provide a sort of internal check for our findings
thereby adding further credibility to the present theoretical "experiments".  

\section{The TD Calculations}
In the  $^3$He + $^4$He$_2^+(v,j)$ collisional event
 two identical bosons $^4$He(S=0)  forming a diatomic ion
collide with a different isotope (fermion) $^3$He giving rise to
 inelastic ($\rightarrow ^3$He$+^4$He$_2^+(v',j')$) and  reactive 
($\rightarrow ^4$He $+^3$He$^4$He$^+(v',j')$)
processes.  

\begin{figure}
  \begin{center}
    \includegraphics[width=0.3\textwidth]{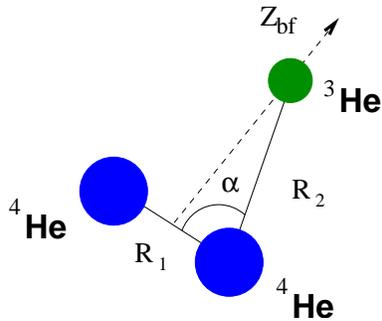}
  \end{center}
  \caption{Coordinates used in the TD code}
  \label{coord}
\end{figure}

\begin{figure}
  \begin{center}
    \includegraphics[width=0.8\textwidth]{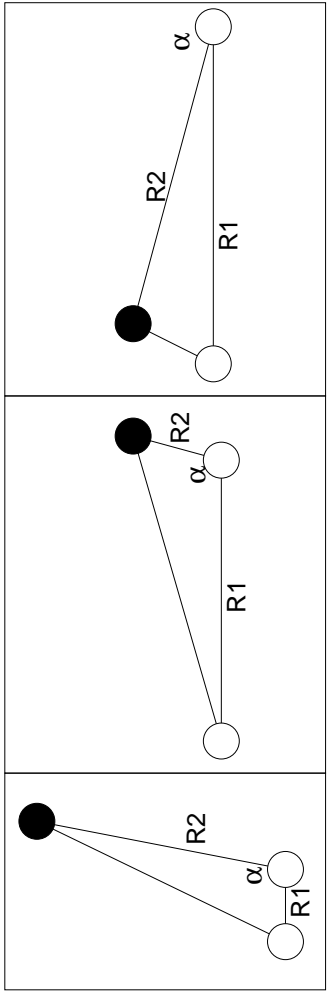}
  \end{center}
   \begin{center}
    \includegraphics[width=0.8\textwidth]{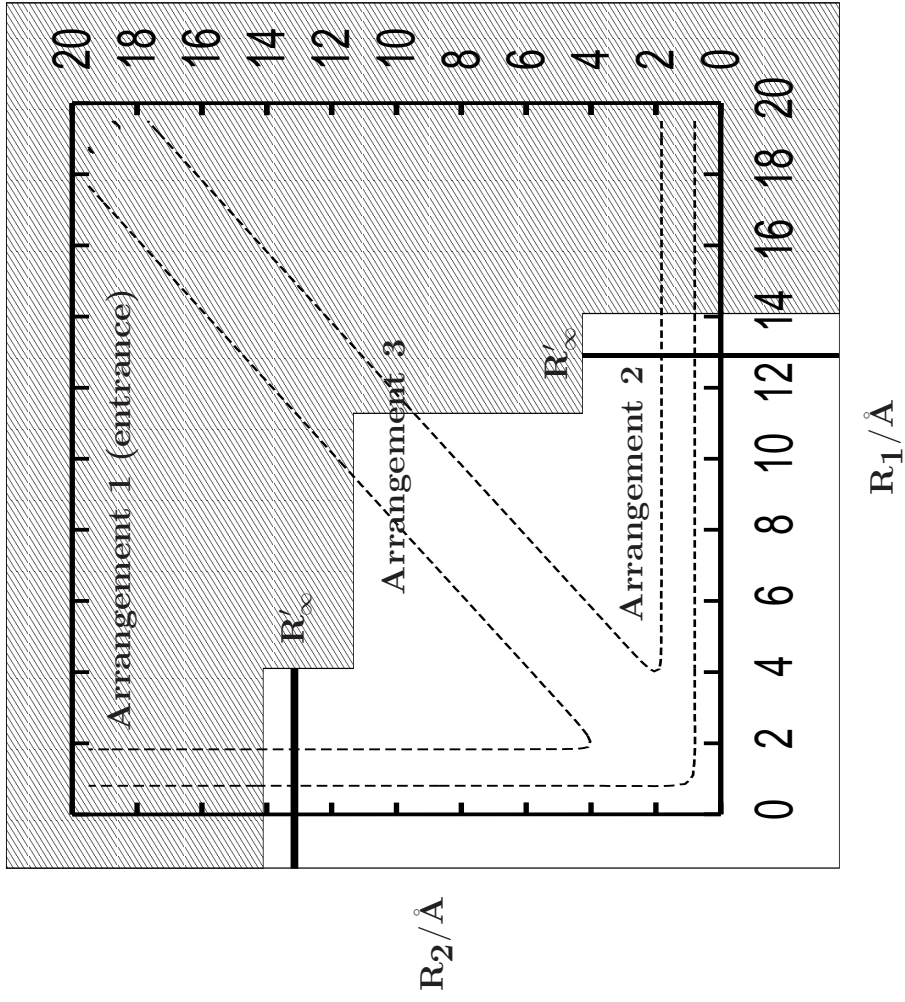}
   \end{center}
  \caption{Possible arrangements in the He$_3^+$
  (upper panel) system and their appearance on the PES of the complex
  (lower panel). See main text for details.}
  \label{arrang}
\end{figure}

In a Body-Fixed (BF) frame of reference defined with respect to the
Laboratory frame by the three Euler angles
$\theta$, $\phi$, $\chi$, we have chosen to use
bond coordinates as the three internal coordinates describing the
relative position of the three particles (see Figure 1) that is two atomic distances
$R_1$, $R_2$ and the angle $\alpha$ between them. 
A method based on these coordinates was presented previously and applied to the 
calculation of state-to-state reaction probabilities for the benchmark collision
$\mathrm{Li}+\mathrm{HF}$\cite{add1}. It was shown there that its use may
present certain advantages over the  more commonly used Jacobi coordinates
approach, for example when dealing with insertion reactions, thus making the bond
coordinate approach method a viable and competitive alternative.One of the
bottlenecks, in terms of computational costs of TD methods, is
represented by the calculation of state-to-state reaction probabilities
because they  would require relatively complicated changes of coordinates. Our
present choice of bond coordinates allows a relatively simple and fast
evaluation of these state-resolved probabilities when only two
molecular arrangements are
accessible of the three which exist in a generic ABC system.
The method is applied here for the first time to a case in which all the
three arrangements are open but where we can still
exploit  the AB$_2$ symmetry for the calculation of state-to-state
probabilities, as we shall further explain later in this section.

A convenient BF frame is the one defined such that the $z$-axis lies along the
vector joining the center of mass of the initial diatom to the
$^3\mathrm{He}$ atom (see Fig. 1) and the three atoms are in the $xz$ plane. 
In the chosen set of coordinates ($\theta$, $\phi$, $\chi$,
$R_1$,  $R_2$, $\alpha$), the complete Hamiltonian for an arbitrary value
of the total angular momentum $J$ takes the form\cite{add1}
\begin{equation}\label{eq:Hamiltonian}
\begin{split}
H=-{\hbar^2\over{2\mu_1}}
   \left(
          {1\over{R_1^2}}{\partial\over\partial{R_1}}
           R_1^2 {\partial\over\partial{R_1}}
   \right)
-{\hbar^2\over{2\mu_2}}
   \left(
          {1\over{R_2^2}}{\partial\over\partial{R_2}}
          R_2^2 {\partial\over\partial{R_2}}
   \right)
+ \left({1\over{2\mu_1 R_1^2}}+{1\over{2\mu_2R_2^2}}\right)
{\bf{\hat L}}^2 +\\
+  {({\bf{\hat J}}^2-2{\bf{\hat J}}\cdot{\bf{\hat j}})\over {2\mu
R^2}} + {\hat T}_{12}  +V(R_1,R_2,\alpha)
\end{split}
\end{equation}
where we denote by $m_0$, $m_1$ and $m_2$ the nuclear masses, with $m_0$ being
that of the "reference" atom, hence $\mu_1=m_0m_1/(m_0+m_1)$,
$\mu_2=m_0m_2/(m_0+m_2)$ are the
reduced masses associated with ${\bf R_1}$ and ${\bf R_2}$, while
$\mu=m_2(m_0+m_1)/(m_0+m_1+m_2)$ is the reduced mass associated with
the ${\bf R}$ Jacobi vector.

In Eq.(\ref{eq:Hamiltonian})
${\bf{\hat L}}^2$ is the angular momentum operator given by the following
expression:
\begin{eqnarray}
{\bf{\hat L}}^2 &=& -\hbar^2\left\lbrace {1\over \sin\alpha}
{\partial\over\partial\alpha}\sin\alpha{\partial\over\partial\alpha}
+{1\over\sin^2\alpha}{\partial^2\over\partial\chi^2} \right\rbrace .
\end{eqnarray}
while ${\bf{\hat J}}$ is the total angular momentum operator.
The term $\left({\bf{\hat J}}^2-2{\bf {\hat J}}\cdot{\bf{\hat
j}}\right) / 2\mu R^2$ where ${\bf{\hat j}}$ is the angular momentum
of the diatomic molecule in the reactants'
Jacobi coordinates, incorporates the
Coriolis couplings given by:
\begin{equation}\label{coriolis}
  \begin{split}
    {\bf{\hat  J}}\cdot{\bf{\hat  j}}= -i\hbar \left\lbrace
    \left({qR_1\over{R_2\sin\alpha}}-\cot\alpha \right)
         {\hat J}_x{\partial\over\partial\chi}
	 -q R_1\sin\alpha {\hat J}_y {\partial\over\partial{R_2}}\right. + \\
	 \left. +\left(1-{qR_1\cos\alpha\over{R_2}}\right)
         {\hat J}_y{\partial\over\partial\alpha}
	 +{\hat J}_z {\partial\over\partial\chi}
	 \right\rbrace
  \end{split}
\end{equation}
where $q=\mu_1/m_0$.

Finally, in Eq. (\ref{eq:Hamiltonian})  ${\hat T}_{12}$ is the
kinetic energy cross term due to the non-Jacobian character of the
internal coordinates we have chosen and has the form
\begin{equation}
  \begin{split}
{\hat T}_{12}={\hbar^2\over{m_0}} \left\lbrace
{\sin\alpha\over{R_1R_2}}{\partial\over\partial\alpha}
-\cos\alpha{\partial^2\over\partial{R_{1}\partial{R_2}}}
+{\sin\alpha\over{R_2}}
{\partial^2\over\partial{R_1}\partial\alpha}
+{\sin\alpha\over{R_1}}
{\partial^2\over\partial{R_2}\partial\alpha} \right \rbrace +\\
-{\cos\alpha\over{m_0R_1R_2}}{\bf{\hat L}}^2
  \end{split}
\end{equation}

When using the bond coordinates we can distinguish three
regions in the PES associated with three different arrangements as described in figure \ref{arrang}: 
the entrance arrangement (that we shall call the channel 1) {[$^4$He$^4$He]$^+$ + $^3$He} and the two other possible
product arrangements (2 and 3) that describe the {[$^4$He$^3$He]$^+$ +
$^4$He} and {[$^3$He$^4$He]$^+$ + $^4$He} combinations. The arrangements 1
and 2 correspond respectively to coordinates $R_1$ or $R_2$ remaining finite while
arrangement 3 would lie along the direction $\alpha=0$ as both $R_1$
and $R_2$ become very large.  The inelastic process will be associated
with the species remaining in channel 1, while both channels 2 and 3
will correspond to  the reactive process: although, apparently
treated in a very different way due to our system of coordinates, these
two channels are physically equivalent.

\begin{figure}
  \begin{center}
    \includegraphics[width=0.4\textwidth]{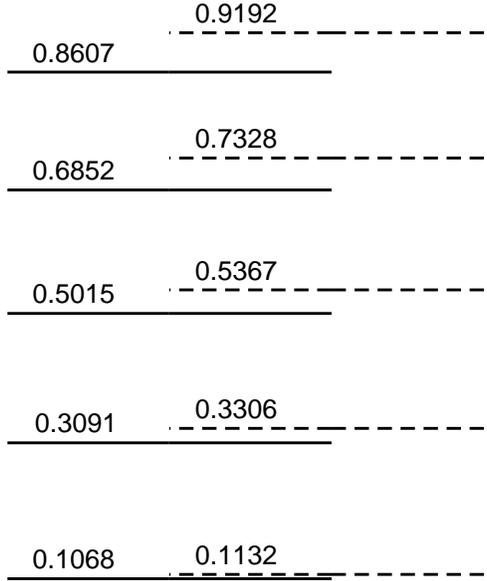}
  \end{center}
  \caption{Lowest ro-vibrational levels of the $^4$He-$^4$He (left) and
  $^3$He-$^4$He(right) molecules. For the former we have reported only
  the $j=1$ levels while for the latter we include only the $j=0$. Energies are in eV, measured
  for the bottom of the entrance valley of reaction}
  \label{channels}
\end{figure}

\subsection{Wavepacket representation}

The total wavepacket is, in the general case, expanded as\cite{add1}
\begin{eqnarray}\label{total-wvp}
 \Psi^{JM \epsilon} (\theta,\phi,\chi,R_1,R_2,\alpha,t)=
\sum_{\Omega\ge 0}^J\quad W^{J
\epsilon}_{M\Omega}(\phi,\theta,\chi)\quad {\Phi^{JM
\epsilon}_{\Omega}(R_1,R_2,\alpha,t)\over R_1 R_2}
\end{eqnarray}
where  $W^{J \epsilon}_{M\Omega}$  are linear combinations of Wigner
rotation matrices\cite{13} of a given parity under inversion
of all coordinates, $ \epsilon$, and where
$M$ and $\Omega$ are the projections of the
total angular momentum, $J$, on the space-fixed (SF) and body fixed (BF)
$z$-axis, respectively.
Insertion of Eq.(\ref{total-wvp}) into the time-dependent
Schr{\"o}dinger equation using the Hamiltonian of eq.
(\ref{eq:Hamiltonian}), generally yields a set of first order
differential equations for the $\Phi^{JM
\epsilon}_{\Omega}(R_1,R_2,\alpha,t)$ coefficients. For the case of
$J=0$ which will be considered in the present calculations the previous
expansion contains only one term. 
The wavefunction is then  given by the solution of the
equation

\begin{equation}\label{eq:difer}
i\hbar{\partial  \Phi \over \partial t} = \left\lbrace
-{\hbar^2\over 2 \mu_1}{\partial^2\over\partial R_1^2}
-{\hbar^2\over 2 \mu_2}{\partial^2\over\partial R_2^2}\right. \nonumber \\
+ \left({1\over{2\mu_1 R_1^2}}+{1\over{2\mu_2 R_2^2}} \right)
     {\bf{\hat l}}^2+ \left. {\hat t}_{12} +  V \right\rbrace \Phi
\end{equation}
with the following meaning of $\mathbf{l^2}$ and  $\hat t _{12}$
\begin{equation}
  \begin{split}
    {\bf{\hat l}}^2 = -\hbar^2 \left({1\over\sin\alpha}
    {\partial\over\partial\alpha}\sin\alpha
    {\partial\over\partial\alpha}\right)\\
    {\hat t}_{12}= \frac{\hbar^2}{m_0} \left[ sin\alpha
    \frac{\partial}{\partial\alpha}\left(
    \frac{1}{R_1}\frac{\partial}{\partial R_2}+
    \frac{1}{R_2}\frac{\partial}{\partial R_1}+
    \frac{1}{R_1R_2} \right) \right.\\
    + \left. \cos{\alpha}\left({1\over{R_1}}
    {\partial\over\partial{R_2}}+
    {1\over{R_2}}{\partial\over\partial{R_1}}
    -{1\over{R_1R_2}}-{\partial^2\over\partial{R_{1}}\partial{R_2}}
    \right) \right] -{\cos\alpha\over{m_0R_1R_2}}{\bf{\hat l}}^2
 \end{split} 
\end{equation}
The integration of the above equation was performed by using the
Chebyshev method\cite{13b} and
 the $ \Phi(R_1,R_2,\alpha,t)$ coefficient is represented
on finite grids for the internal coordinates $R_1$, $R_2$, $\alpha$.
A set of equidistant points, $R_1^i, R_2^j$, was chosen for the
rectangular  bidimensional radial grid ($n_1 \times n_2$), which
allows the evaluation of the radial kinetic terms using Fast Fourier
Transforms (FFT)\cite{13c}. For the angle $\alpha$ a set of
$n_{\alpha}$ Gauss-Legendre quadrature points, $\alpha^k$ (with
weights $\omega^k$), is used.  Thus,  the terms  involving
derivatives in $\alpha$ are evaluated through a discrete variable
representation (DVR) transformation which reduces the
procedure to  a simple multiplication of a matrix by a vector
\cite{add2,add3,add4,add5,add6,add7}.
 The grid representation of the
wavepacket is then given by
\begin{eqnarray}
[\Phi]_{ijk}=\Phi(R_{1}^i,R_{2}^j,\alpha^{k}) \sqrt{\omega^k}
\end{eqnarray}
where, for convenience, the Gauss-Legendre weights are introduced.

In order to use a finite bidimensional radial grid,
 the wavepacket is absorbed at each
time step by multiplying the wavepacket for $f_1(R_1) f_2(R_2)$,
where $f_i(R_i)= exp\lbrack-\Upsilon_i(R_i-R_i^{abs})^2\rbrack$ for
$R_i>R_i^{abs}$ and $f_i(R_i)= 1$ otherwise. The absorbing regions are
presented in the  lower panel of
Figure \ref{arrang} as a shaded area.

\begin{table}
\caption[table1]{ Parameters used in the wavepacket propagations
for $\nu=0,\,1,\,2\,\,\,j=1\,\,\mathrm{and}\,\,J=0$; distances in \AA, times in
ps. The gaussian parameters $ ({\cal K}_0, \Gamma)$ in Eq (9), are given by
$(14.30,\,
0.42),\,(10.87, 0.55),\,(6.27, 0.96) $ respectively }
\begin{center}
\begin{tabular}{|c|c|}
\hline 
 ($R_1^{min}(\AA),R_1^{max}(\AA),n_1) $  & 0.69,\quad  19.50,\quad  360\\
\hline   
 ($R_2^{min}(\AA),R_2^{max}(\AA),n_2) $  &  0.69,\quad  19.50,\quad  360\\
\hline   
 $n_{\alpha}$ & 130   \\ 
\hline  
 ($R_{1}^{abs}(\AA), \Upsilon_1$)  & 14.10,\quad  0.016 \\ 
\hline  
($R_{2}^{abs}(\AA), \Upsilon_2$ ) & 14.10,\quad  0.016 \\ 
\hline  
 ($R_0(\AA),R'_{\infty}(\AA)$)  & 13.00,\quad  13.00\\ 
\hline  
 $\Delta t(ps)$ & 0.003 \\ 
\hline
\end{tabular}
\end{center}
\end{table}
The actual parameters of the propagation used in the calculations are listed in
Table I. It should be noted that the number of angles, $n_{\alpha}$, required to
converge is indeed very large: channel 3 is asimptotically open
only for
$\alpha=0$ and many angular quadrature points are required in order to have a
grid dense enough in the small-angle region.
The initial wavepacket represents
the reagents approaching the
collision region with the diatom in a given $(\nu,j)$ ro-vibrational state
and with a continuum distribution of relative kinetic energies.
The initial wave packet can thus be  expressed  (in reactants' Jacobi coordinates
$(r,R,\gamma)$) as a product of a  diatomic wavefunction  and a Gaussian function for the
relative translational coordinate:
\begin{eqnarray}\label{initialwp}
\Phi(t=0)=\chi_{\nu j}(r)Y_{j\Omega_0}(\gamma,0)G(R)
\end{eqnarray}
where $G(R)$ is a complex Gaussian function written as follows
\begin{eqnarray}
G(R) = \left({2\over \pi \Gamma^2}\right)^{1/4} \exp\left\lbrack
-{(R-R_0)^2\over \Gamma^2} - i {\cal K}_0 (R-R_0)\right\rbrack.
\end{eqnarray}
The gaussian is centered at  a convenient $R_0$ value such that
the interaction between the reactants can be considered negligible. The energy
distribution for the initial wavepacket $a(E)$ is then determined
using
\begin{eqnarray}\label{eq:weightE}
a(E)=\left(\frac{\mu}{2 \pi \hbar^2 k_{\nu j}} \right)^{1/2}\int
e^{ik_{\nu j}R} G(R)dR
\end{eqnarray}
with , $k_{\nu j}=\sqrt{2\mu(E-E_{\nu j})/\hbar^2}$.

This system contains two identical bosons and hence the total
wavefunction must obey the correct spin statistics: the total
wavefunction must have the correct symmetry under the action of the
$P_{12}$ operator that exchanges the two bosons. This means that (as
in all the  AB$_2$ systems) the calculations with even and odd initial
$j$ values for the diatomic wavefunction are decoupled: since the
Hamiltonian commutes with the parity operator $P_{12}$, a wavepacket
built as an eigenfunction of $P_{12}$ (as it occurs with our initial wavepacket
$\Phi(t=0)$) will remain so under time evolution.
An additional simplification in our calculations stems from the
consideration that, since the $^4$He$^+_2$ is a $\Sigma^+_u$ molecule composed
of two spinless nuclei, only odd rotational states are allowed in the entrance
arrangement. As a consequence of these constraints, it is clear that the
appearance of the forbidden even-j rotational states in the inelastic
distributions obtained from the evolved wavepacket using eq. (11) below would
indicate convergence problems. This was one of the various criteria we used for
checking convergence in our present calculation and we found no instance where
such states would appear.

The main steps of the TD method applied to the J=0 case were the
following \cite{add1}:
\begin{itemize}
\item The initial wavepacket, obtained in the reactants' Jacobi coordinates $(r,R,\gamma)$
is transformed into bond coordinates $(R_1,R_2,\alpha)$;
\item The propagation is performed using bond coordinates;
\item The S matrix elements for the transition into any other reactive or inelastic final state
can be obtained from the evolved wavepacket  using the product
asymptotic analysis method of Balint-Kurti {\em et al.} \cite{15,15b} 
that requires a change of coordinates to the Jacobi system
$(R'_\infty,\,r',\,\gamma')$ corresponding to the final arrangement. 
From the transformed wavepacket at each time step,  the state-to-state
probabilities are obtained using the expression
\begin{equation}\label{acidez}
  P_{\nu j \rightarrow \nu'j' }(E)=
\frac{1}{|a(E)|^2}\frac{k_{\nu'j'}}{2\pi \mu}
|A_{\nu'j'}(R_\infty ' ,E)|^2
\end{equation}
where, as noted above,  $k_{\nu'j'}=\sqrt{2\mu (E-E_{\nu'j'})/\hbar^2}$,
$\mu$ is the reduced mass of the desired  arrangement
channel, $R_\infty '$  is a value of the Jacobi radial
coordinate well into the asymptotic region and the
$A_{\nu'j'}(R_\infty ',E)$ are the energy transforms defined for the desired
channel as:
\begin{equation}\label{eq:zCvj}
A_{\nu'j'}(R_\infty ',E)=  \int_0^\infty dt e^{iEt/\hbar} \langle
\chi_{\nu'j'}(r') P_{j' 0}(\cos{\gamma} ') \vert \Phi(R_{\infty}',r
',\gamma ',t)\rangle
\end{equation}
\end{itemize}

The computational cost of the required coordinate transformations is not very
high because the arrangement channels 1 and 2 share one of the Jacobi radial
coordinates with the bond coordinates system $r'=R_i$, thereby reducing
the numerical effort due to coordinates tranformations: even in the general
case of $J \ne 0$, only simple rotations on the plane identified by the three
atoms would be required to move to the desired BF system. Arrangement
3 would require a much more expensive change of coordinates, but due to the
symmetry, the probability for the $^3\mathrm{He}$ to strip either of the two $^4\mathrm{He}$ atoms
must be exactly the same. Hence, we can assume that the probabilities in channel 3 are
the same as those in channel 2 and thus we can avoid calculating them; any deviation from unity
of the sum of all of them can then be taken as a sign of lack of convergence.
This has been another criteria in determining the convergence of our
calculations in the actual situations of this work.

\subsection{The TI Calculations} 

We have employed  the \emph{abc} program of
Skouteris et al. \cite{manolopoulos00}
which performs an expansion of the total wavefunction in Delves hyperspherical
coordinates \cite{pack87} and uses the coupled channel method formulated
by Schatz \cite{schatz88} with an additional orthogonalization scheme
of the vibrational basis at fixed hyperradius to avoid over-completeness
of the basis set in the short range region \cite{parker93}. The coupled
channels equations are then solved using the constant reference potential
log-derivative algorithm of Manolopoulos \cite{manolopoulos86}.

The convergence of the
resulting S matrix is sensitive to two parameters, the hyperradius
\( \rho _{max} \) at which the asymptotic condition is imposed and
the  step-size \( \Delta \rho  \) of the propagator. Converged results
within 1\% were obtained with \( \Delta \rho  \)=0.02 a.u. and \( \rho _{max}=25.0 \)
a.u. for the lowest collision energy considered here ($\sim$7 meV). 

The basis set was built using all the ro-vibrational levels of the
two [He-He]$^+$ diatomics  whose energies lie below
the cut-off energy value of 1.5 eV as measured from the bottom of
the entrance channel. Since our calculations would
have otherwise involved the use
of too many rotational levels, we have imposed a maximum value of \( j_{max} \)=20 for
each molecule which is enough to obtain probabilities converged within
5\% also for the highest rotational states included. The basis set chosen has
been able to reproduce almost perfectly the TD results also at the higest
energies employed here as we shall see below.  
Our basis set comprised of
a total of 235 basis
functions for $J=0$. As noticed before the
[$^4$He-$^4$He]$^+$ molecule can only have odd rotational states. 

We have performed the calculations for about a thousand different collision
energies on a grid ranging from 7.375  meV up to 1 eV. The
lowest collision energy has been chosen to be 1 meV higher than the
energy necessary to make  the reactive channel for the title reaction
to be energetically accessible
when considered in the following direction
\begin{equation}\label{eq:react}
  \mathrm{^4He}_2^+ + \mathrm{^3He} 
  \longrightarrow \mathrm{^3He^4He}^+ + \mathrm{^4He}
\end{equation}

The small endothermicity of 6.375 meV is due to the difference of zero
point energy between the $\nu=0,j=1$ state of the reactants in eq. (\ref{eq:react}) and
the  $\nu=0,j=0$  state of the heteronuclear product molecule.
For clarity we report in Figure \ref{channels} a diagram of the
vibrational energy levels involved in the calculations.

\section{Results and Discussion}

\begin{figure}
  \begin{center}
    \includegraphics[width=1.0\textwidth]{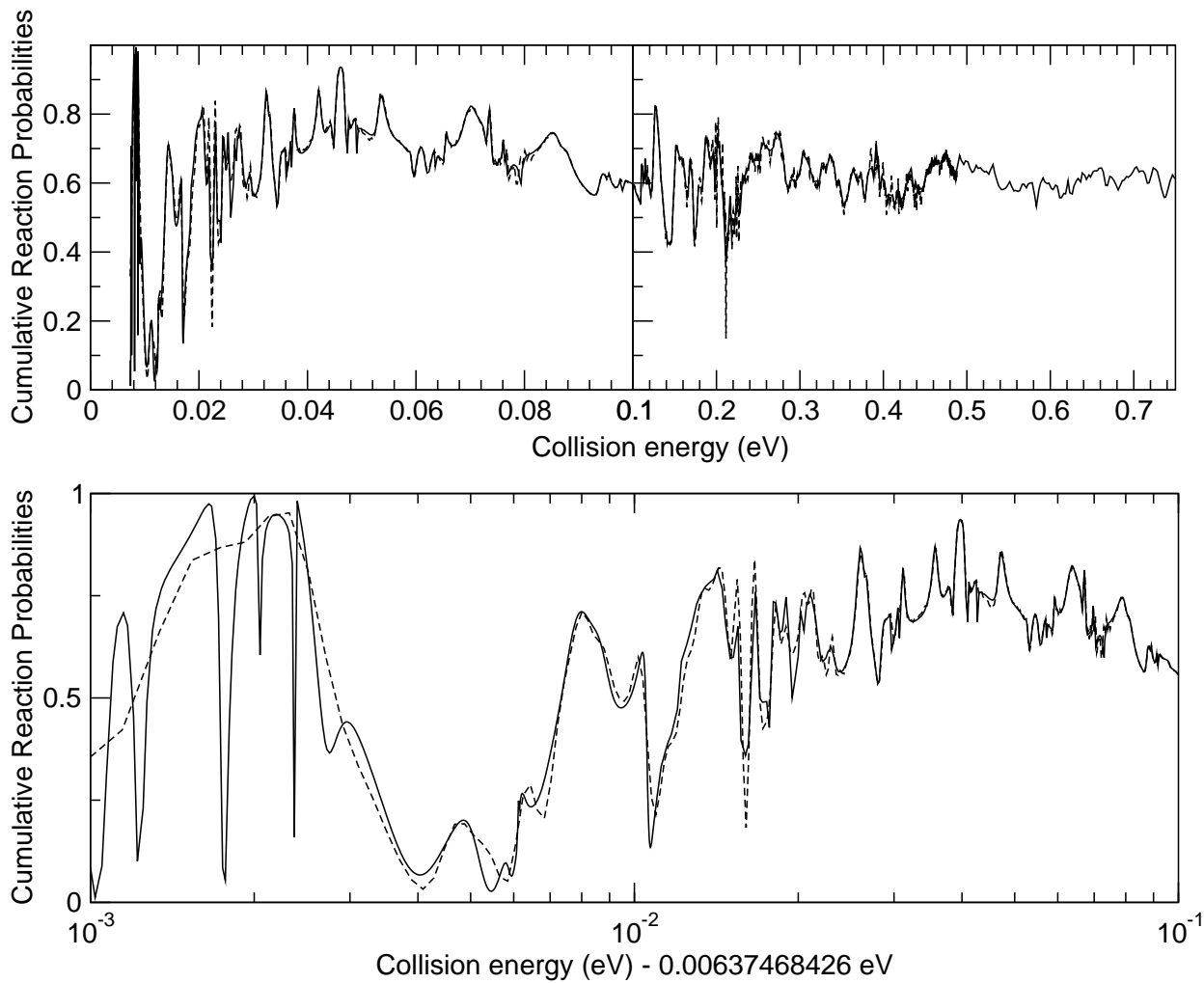}
  \end{center}
  \caption{Reaction probabilities for TD (dotted lines) and TI
(solid lines) calculations for
the initial state $(\nu=0,\,j=0)$. 
    The lower panel shows the same probabilities on a logarithmic scale for
the lowest energies above reaction threshold}
  \label{comparison}
\end{figure}

In Figure \ref{comparison} we report the reaction probability when
starting with the
initial state $(\nu=0,j=1)$ and summing over all the final accessible states.
The two sets of data that refer to the two different, independent
calculations we have mentioned above are in good agreement over the whole
energy region of interest  and provide a useful internal check
of our calculations. The small differences are due to different causes: first
the truncated size of our basis set expansion in the TI code is the source of a
small level of inaccuracy especially for the transitions involving the higher
rovibrational states; in second place, the TD calculation 
become inaccurate (as expected in general) for very low kinetic energies (below
10 meV in the present case) because of the finite time for progation that
necessarily neglects the contribution to the correlation functions coming from
longer times. 
The \emph{abc} program uses in fact a basis set
which is constructed as an orthogonalized product of ro-vibrational arrangement
wavefunctions. As a consequence, the description of reactions where
triangular geometries are dominant (e.g. Li$_2$+Li system) may be
inaccurate. Although our system here has clearly shown a transition state
and a minimum energy path that is mainly collinear \cite{7,8}, we
still intend to provide a consistency check of our calculations by comparing 
the TI results with those from a TD method
that is not based on basis sets expansion.  On the other hand, the TD
method is based on a time propagation in bond coordinates and might
therefore present some
problems when applied to a A-B$_2$ system where the third arrangement
(channel 3 in Figure \ref{arrang}) becomes energetically accessible when
$\alpha \sim 0$; this would require many angular 
points in order to  describe the reactive process as discussed earlier on in
this work. 

\begin{figure}
  \begin{center}
    \includegraphics[width=1.0\textwidth]{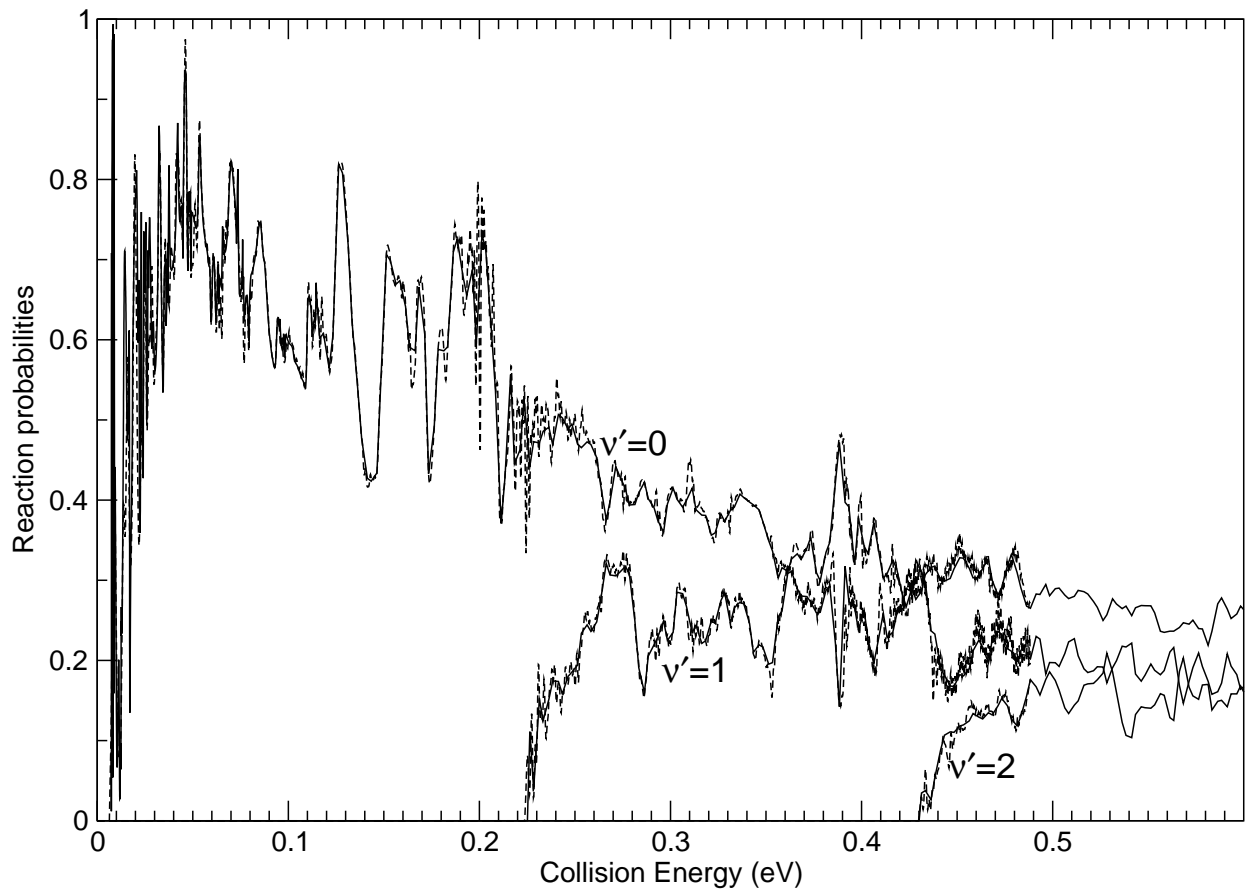}
  \end{center}
  \caption{TI (solid lines) and TD (dashed lines) rotationally summed reaction
probabilities for different final
vibrational states. The initial state of the reactant molecule is
$(\nu=0,\,j=0)$.}
\label{crp-ground}
\end{figure}

The pattern of resonances that can be seen in the reaction
probabilities profile is very complicated as many of these
features may  be due to the opening of the various
asymptotic channels of the two diatomics involved in the reaction. Many
others are likely to be due to Feshbach-type resonances with metastable states
of the triatom and
to potential trapping due to centrifugal barriers in the exit channels, but it
would be premature to analyze them now in any detail given the
absence of some experimental indication as to their physical presence. 

In Figure \ref{crp-ground} we report the individual contributions of the
reaction probabilities for different final vibrational channels (the
probabilities are rotationally
summed). As can be seen from that figure, the total reaction probability is
larger than 60\%
over most of the range of collision energies considered here (the
collision energy in this figure is measured with respect to the
$^4$He$_2$($\nu=0,j=1$) channel i.e. to the reactants' ground
state). However, at large collision energies the probability of
populating the ground vibrational state of the products decreases and
the flux is more or less equally redistributed over all the open, final
vibrational states.

In the highly symmetrical systems like  A$_3$  one should expect that
the total reaction probabilities would be exactly the same as the
non-reactive one simply because the two fluxes are not
distinguishable. In our case, however, a certain degree of
asymmetry is introduced by the isotopic change and therefore  the
reactive process becomes more likely to occur than the inelastic collision
process. This feature, at high energies (above 1.0 eV) is however mainly due to
the fact that the there are two arrangements containing the product AB
molecule. At these energies, in fact, the helium exchange reaction has a
total probability of roughly 60\%, 30\% for each of the two possible exchange,
while we find a 40\% probability of simple elastic/inelastic non-reactive
scattering. This behavior may be attributed to the mass difference because it
seems reasonable that when the collision takes place at high energies the
lighter $^3$He may be not efficient in substituing the heavier isotope.
We also believe that, given the relatively high collision energies
employed here, the small difference of zero point energy mentioned in
the previous section has a very small effect on the dynamical
behavior.
Below 1.0 eV we see from Figure \ref{comparison} that the probability may rise
well above 0.6 and reaches the range 0.8-1. In this case the probability of
a non-reactive scattering is less than 20\%. We believe that the dominant
effect at these lower energies may be the higher density of states
of the product AB molecule with respect to the reactant
A$_2$ molecule which has only odd rotational states for the symmetry reasons
mentioned before.

\begin{figure}
  \begin{center}
    \includegraphics[width=1.0\textwidth]{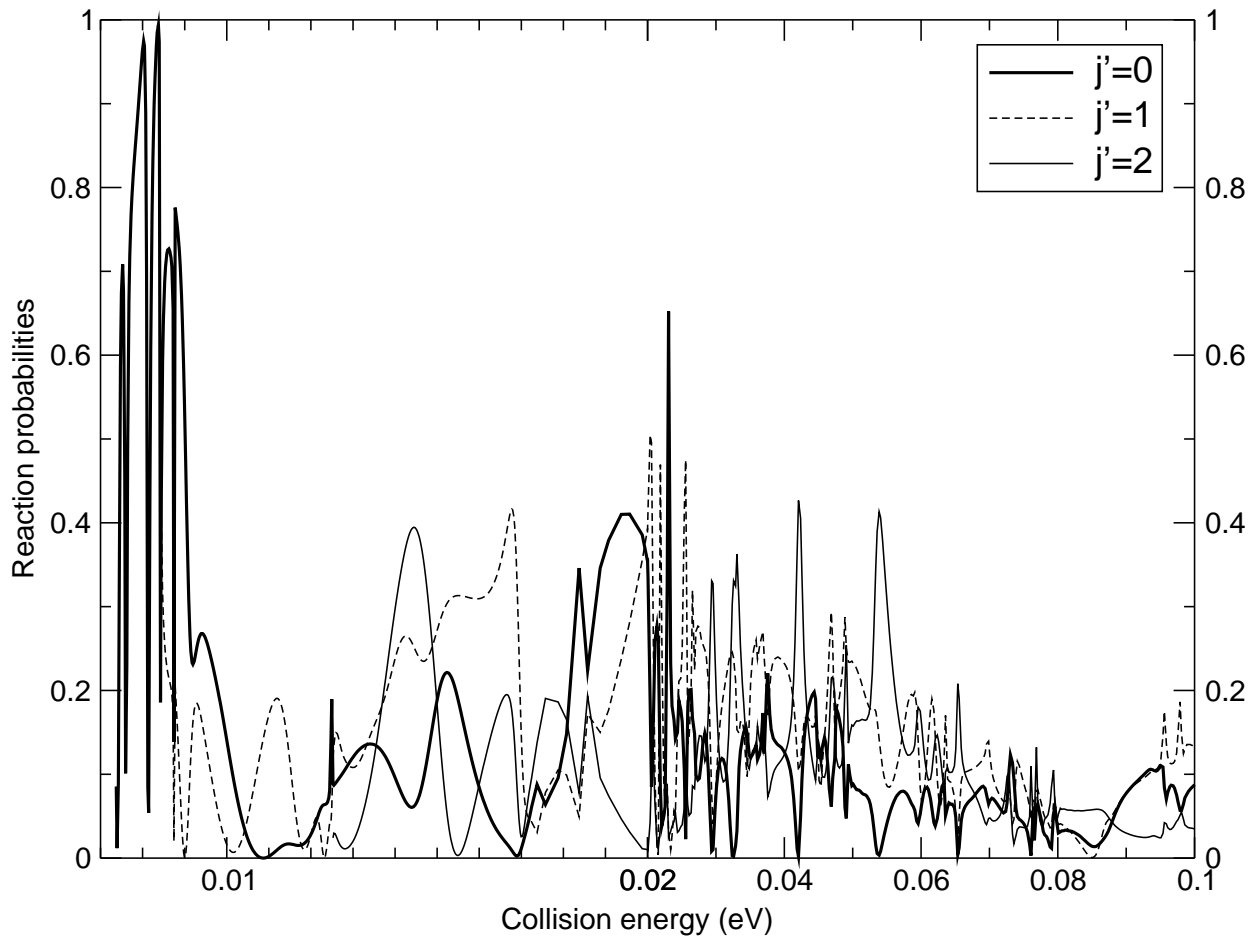}
  \end{center}
  \caption{A limited portion of the TI state-to-state reaction probabilities for
ground state reactants and various final rotational states i.e. for the reaction
$\mathrm{^4He}_2^+(\nu=0,\,j=0) +\mathrm{^3He} \rightarrow
\mathrm{{^4He}{^3He}}^+(\nu=0,\,j') +\mathrm{^4He}$ where the final molecule is
always in the $\nu=0$ vibrational level}
  \label{rot}
\end{figure}

Much more difficult is the task of finding significant patterns of
behavior in the final products' rotational distributions. The first
three  state-to-state reaction probabilities for
different final rotational states, when the reaction begins with the ground
state reactants are reported in Figure \ref{rot}. We can
immediately see there that high degree of variation of each individual
probability persists as $j$ changes as it also does for the other open channels
not shown in the Figure. At higher energies (not shown in
the Figure), on the other
hand,  the various final rotational states of the product molecule are
all substantially populated.  

\begin{figure}
  \begin{center}
    \includegraphics[width=1.0\textwidth]{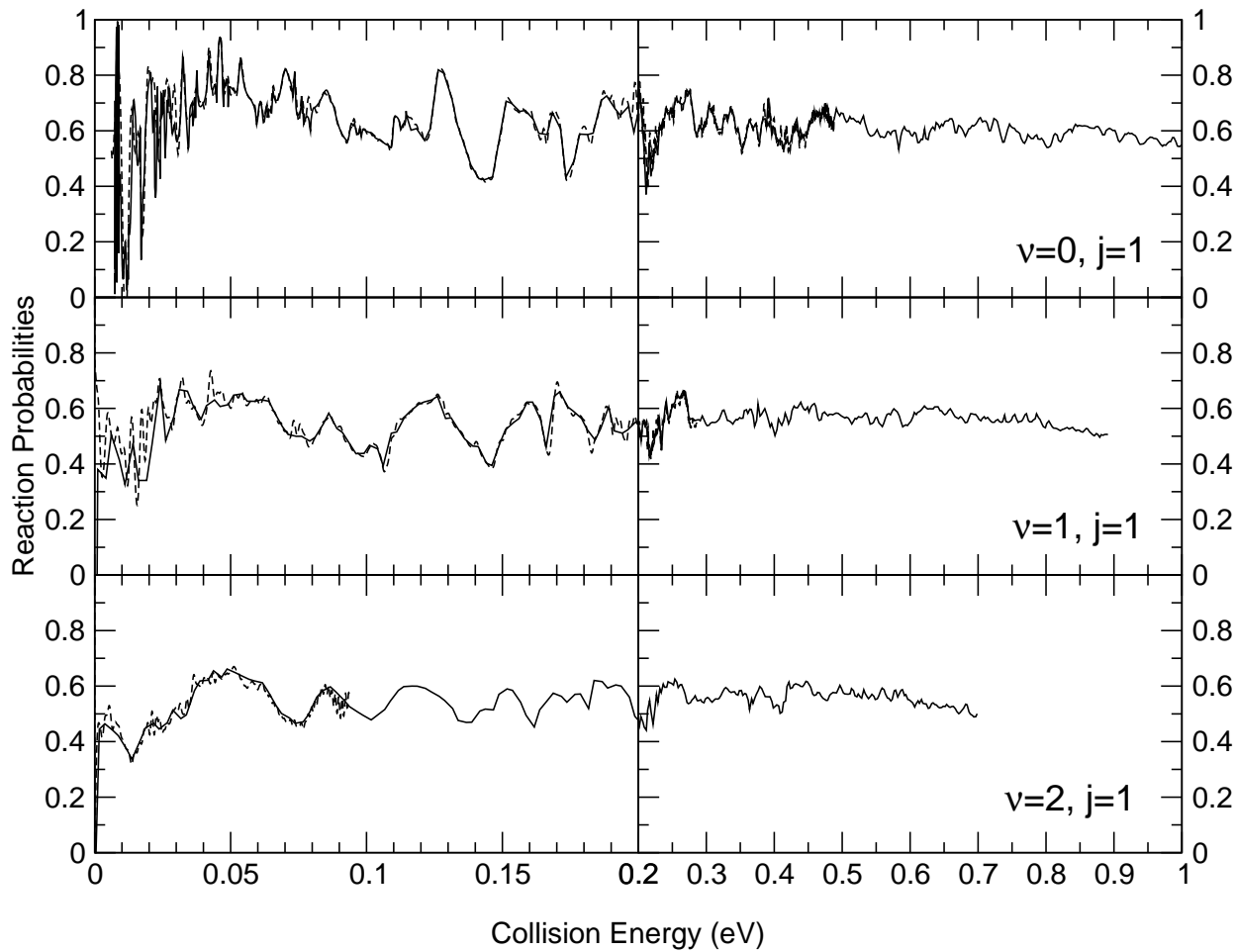}
  \end{center}
  \caption{TI reaction probabilities for different initial
vibrational states (solid lines); TD results are also reported as dashed lines}
  \label{vib}
\end{figure}

Another important piece of information that we may obtain from our
calculations is the effect of increasing the internal energy (i.e. the
ro-vibrational excitation) of the reactants.
In Figure \ref{vib} we therefore report the total reaction probability
summed over all the open final states  but now for three different initial
states $(\nu=0,j=1)$, $(\nu=1,j=1)$
and $(\nu=2,j=1)$ (the collision energies have been selected relative
to the specified initial state).  As it can
be easily seen
for this figure, the initial degree of vibrational excitation (at
least for the lowest states examined here) does not change significantly the
overall dynamical behavior of the system: the reaction probabilities
rise rapidly at threshold from zero to $\sim 60\%$  and remain
roughly of that size when the collision energy increases. 

\begin{figure}
  \begin{center}
    \includegraphics[width=1.0\textwidth]{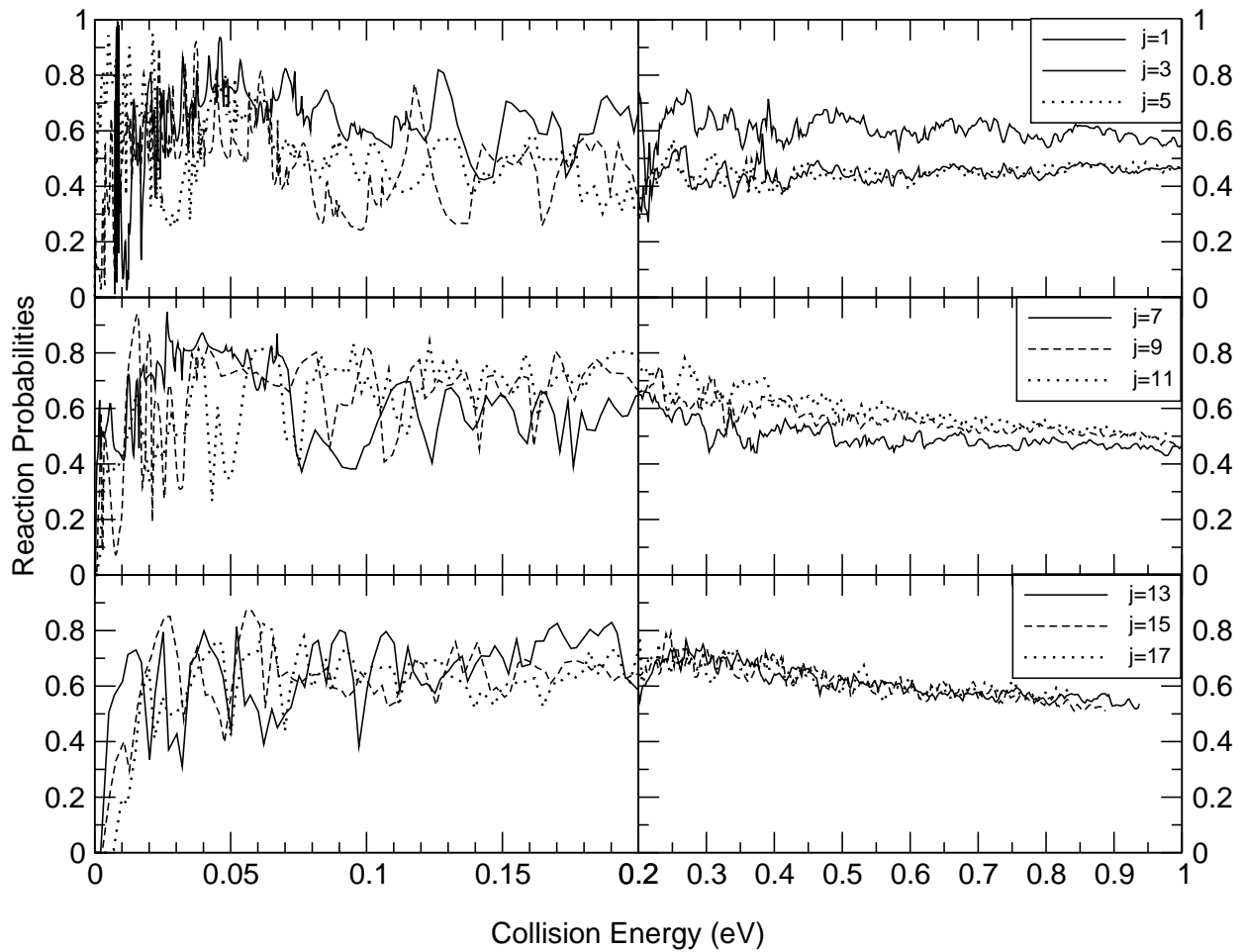}
  \end{center}
  \caption{TI reaction probabilities for different initial rotational
states (the reactant molecule is in the $\nu=0$ vibrational state)}
  \label{rot2}
\end{figure}

A similar situation has been obtained when considering
the reactions probabilities produced by rotationally excited
reactants. For example when looking at the results reported in
Figure \ref{rot2} we see how even a large rotational energy content does not
produce a corresponding increase in reaction probabilities. It is worth 
pointing out however that we are limiting this analysis to the $J=0$ case and
therefore a situation in which the initial molecular angular momentum is
balanced by a corresponding relative orbital angular momentum $l$. This means
that the reaction paths that we sampled here with rotationally excited
molecules would present centrifugal barriers. 
In conclusion, however, we can say that the
reaction dynamics in our system is probably dominated by resonances and
therefore the memory of the initial rotational state is soon lost during the
reaction.

\section{Higher angular momenta}

To increase the value of the total angular momentum beyond $J=0$ for
reactive scattering is still a  computationally demanding task. 
From the analysis of the lower partial waves, however,
we can already extract useful information about the overall dynamics.
When going to the situation with $J \ne 0$ the number of coupled channels increases due
to the proliferation of the possible $\Omega$ values, where $\Omega$ is the
helicity quantum number \cite{manolopoulos00}. Furthermore, for each
$J>0$  value we have two possible values of the total parity
eigenvalues ($p=\pm 1$) which require two independent calculations. The
calculations for $J\ne 0$ that we present here are "exact" (no approximations
such as coupled states have been used) and have been done using the TI method
outlined above and the \emph{abc} code.

\begin{figure}
  \begin{center}
    \includegraphics[width=1.0\textwidth]{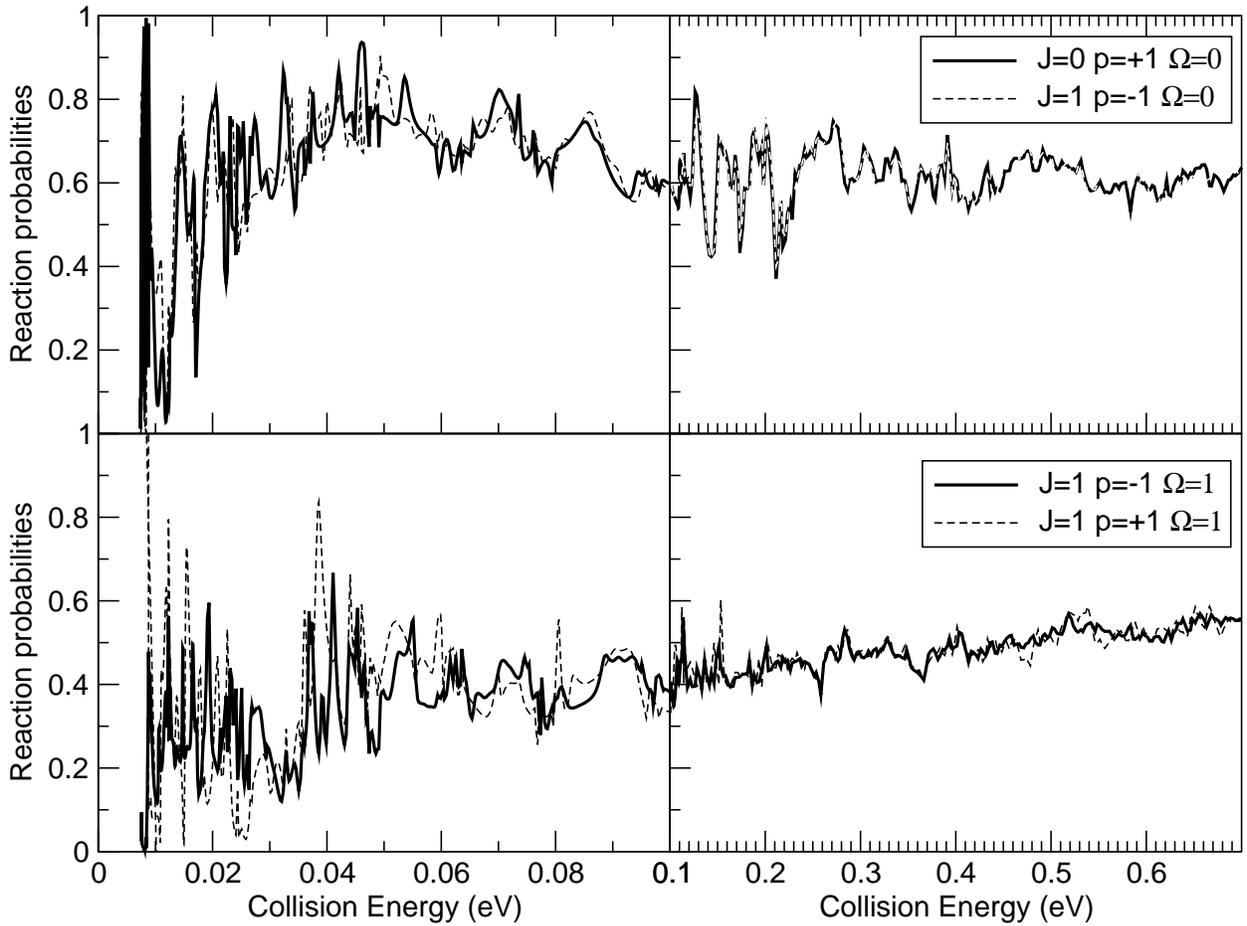}
  \end{center}
  \caption{TI  reaction probabilities for $J=1,\,\Omega=0$ (upper
panel) and for $J=1,\,\Omega=1$ (lower panel)}
\label{j1}
\end{figure}

\begin{figure}
  \begin{center}
    \includegraphics[width=1.0\textwidth]{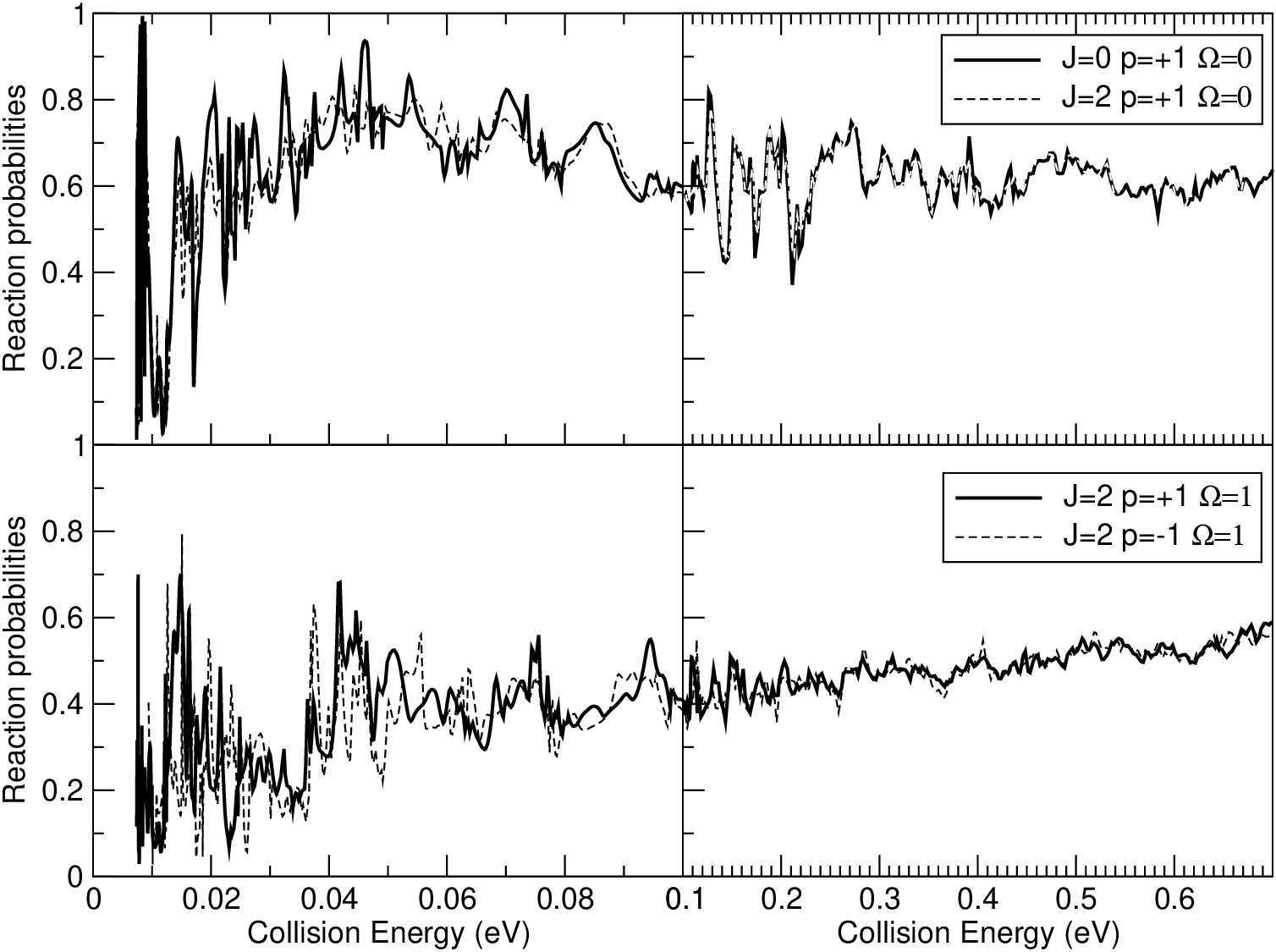}
  \end{center}
  \caption{TI  reaction probabilities for $J=2,\,\Omega=0$ (upper
panel)
and for $J=2,\,\Omega=1$ (lower panel)}
\label{j2}
\end{figure}

We report in Figure  \ref{j1} and \ref{j2} the total  probabilities
for $J=1$ and $J=2$ for the reaction which starts with
$^4$He$_2^+(\nu=0,j=1)$. In both cases the results are compared
with the $J=0$ reaction probabilities (thick solid lines in the upper
panels). As can be seen in the upper panels of those figures the reaction probabilities
for higher angular momenta are very similar to the ones for $J=0$, the only
difference being the small shifts in the resonance pattern. This is true
also at relatively low energies although the resonance pattern is
becoming different especially close to threshold.

When looking at the lower panels of Figs.  \ref{j1} and \ref{j2} we
can further see the contributions due to the initial $\Omega=1$ helicity
component of the
reaction: as it may be expected for a reaction which has a marked collinear
constraint this last contribution is significantly lower than for
$\Omega=0$ especially at lower energies. 

Given the fact that accurate calculations for other values of $J>0$ would
require computational times which are too long and since we are
here in presence of a ionic potential which in principle may require many total
$J$ in order to properly converge to a reactive cross section, we have decided
to calculate the reaction rate constants by using something similar to a
$J$-shifting  approximation, i.e. by using only the $J=0$ reaction
probability $P(E)$ to estimate the total rate constant. A realistic
approximation when dealing with barrierless (ionic) systems like the
one we are examining here, has been suggested in Ref \cite{jshift}: this
approximation essentially consists in obtaining  the $J\ne
0$ reaction probabilities by judiciously ``shifting'' the $J=0$
values. We have here used the
formula: 
\begin{equation}
P_J(E;\nu,j)\sim P_{J=0}(E-V^*_J;\nu,j) 
\end{equation}
where $\nu,j$ identify the initial state of the reactants and $V^*_J$ is
the height of the centrifugal barrier of the entrance channel taken
along  a suitable monodimensional
potential generated by a non-zero initial orbital
angular momentum $l$ value that is allowed for a given $J$ and $j$; 
we have chosen to use the potential of a collinear geometry given by
$\mathrm{He}-r_{eq}-\mathrm{He}-r-\mathrm{He}$ with $r_{eq}=2.046$ a.u.. 
The only difference with ref. \cite{jshift} is that we have initially
a $j=1$ molecule which means that for each $J$ value there are
three allowed values of $l$ given by $|J-1| \le l \le J+1$.  Since
this happens for all $J$ values that may contribute, the further initial-j 
averaging leaves the cross section unchanged 
\begin{equation}
\sigma(E_{coll};\nu,j)=\frac{\pi}{k^2_{coll}}\sum_J (2J+1)
P_J(E;\nu,j) 
\end{equation}

\begin{figure}
  \begin{center}
    \includegraphics[width=1.0\textwidth]{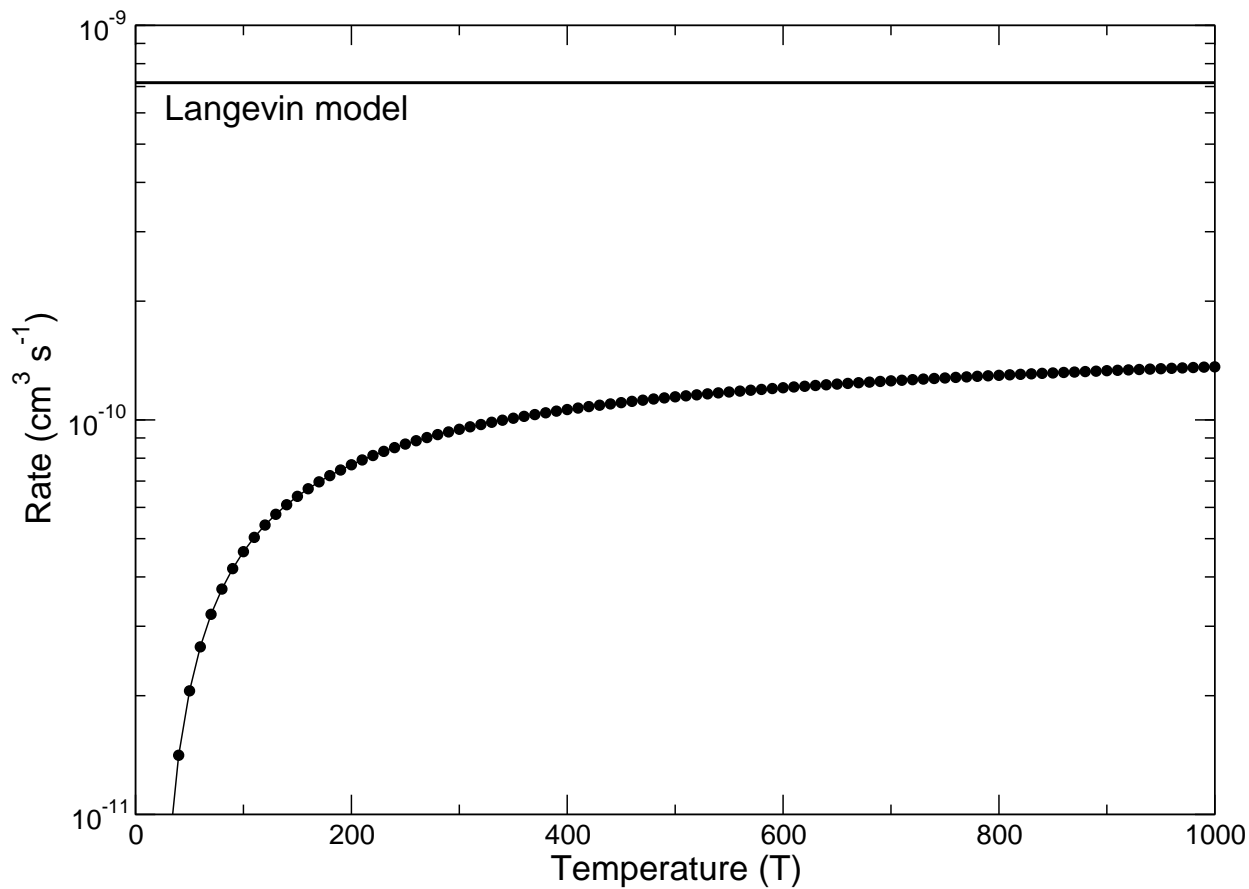}
  \end{center}
  \caption{Rate constants for the reaction {$\mathrm{^4He}_2^+ +
  \mathrm{^3He}  \longrightarrow \mathrm{^3He^4He}^+ + \mathrm{^4He}$}
  as a function of temperature. Also shown as an horizontal line is
  the Langevin value.}
\label{rate}
\end{figure}

From this cross section we finally obtain the rate constants reported in Figure
\ref{rate}. Although this is a very approximate prescription, our
final result can still tell us that the rate constant for a reactive exchange
process in He$_3^+$ is of the order of $10^{-11} - 10^{-9}$
cm$^3\cdot\mathrm{s}^{-1}$ a value that is significantly lower than
the Langevin capture rate also reported in Figure \ref{rate}. One should note
also that, although Langevin capture rates assume the reaction to be exoergic,
the very small endoergicity of the present system still allows us to use it for
an estimate of the reaction rates.

\section{Present Conclusions}

We have presented new theoretical results on the reactive dynamics of He$_3^+$
system. In particular, we have discussed the reactive behavior for the
lowest total angular momenta and have  extended the
calculations to obtain approximate rate constants for the title reaction.  In
order to make  the   process physically
clearer and to get a better  insight into its mechanism, we have substituted one
of  the $^4$He with an $^3$He so that
we have introduced a small endothermicity in the atom exchange
process that is occurring during the reactive event. In this way we
have also  introduced an asymmetry in the density of
states associated with the reactants (here an homonuclear molecule) and the
products (here an heteronuclear molecule). This difference makes for a more
favorable reactive process with respect to to the simpler inelastic collision
process as shown by our calculations. 

The reaction considered here has a collinear MEP without 
activation barrier and therefore behaves in the main like a typical ionic
reaction. We have obtained complete state-to-state probabilities for the
reaction using a time independent method and then, because there may be some
issues related to the use of an arrangement-based basis set expansion, we have
checked the numerical reliability of our TI findings by further reproducing our
results using a time dependent procedure.

Our present results clearly show that the reaction under study turns out to be
quite efficient when $J=0$ and
represents more than  60\% of the scattered flux at the energies
considered.  Internal excitation
of the colliding partner does not appear to produce  substantial increases of
the reaction probabilities at least for the lowest vibrational and rotational
states of the reactants. Even with relatively highly rotationally
excited He$_2^+$ (up to  $j=19$) the size of the reactive probability remains
very similar to the one for the non rotating molecule. This may be due to
the fact that the reaction mecanism is going through many resonances, thereby
dynamically loosing the effects of having "hot" reactants. 

Furthermore, by looking at the final distributions over the vibrational channels
we have seen that the reaction can easily produce vibrationally excited
molecules and the reactive flux is more or less equally redistributed over all
the open, final vibrational states.  The final rotational populations, as far
as we can judge from our data, seem to be strongly dependent on the collision
energy and do not show any simple pattern of interpretation. 

We have also performed fully converged calculations for $J=1$ and
$J=2$ and we have seen that the latter data provide very similar results
to $J=0$ calculation although the resonance patterns slightly change and
shift in energy, as one expects from such systems. We have thus been able to
 provide rate constants for
the reaction through the use of a $l$-shifting approximation \cite{jshift} which
yields final rates of the order of $10^{-10}$
cm$^3\cdot\mathrm{s}^{-1}$. The latter results indicate that in the
present system the reactive process may be as efficient as the
inelastic vibrational de-excitation process estimated earlier by us \cite{7}
in providing a mechanism of energy exchange in the droplets. The size of the
exchange rate also suggest that observation of such reaction may be possible in
the droplet environment.

\begin{acknowledgments}
The financial support of the Scientific Committee of the University of
Rome, of the CASPUR Supercomputing Center and of the INTAS grant n.
03-51-6170 is gratefully acknowledged. M.L. thanks the "cold molecules" TRN
n. HPRN-CT-2002-00290 for supporting his stay in Rome, where this work begun.
The financial support of the same RTN "cold molecules" is also acknowledged.
\end{acknowledgments}

\end{document}